\documentclass[reqno,11pt]{amsart}

\usepackage{amssymb,graphics,graphicx,wrapfig}
\usepackage{epic}

\def\be{\begin{equation}}
\def\ee{\end{equation}}
\def\ba{\begin{align}}
\def\bm{\begin{multline}}
\def\bfig{\begin{figure}[htb]}
\def\efig{\end{figure}}
\numberwithin{equation}{section}
\newtheorem{theorem}{Theorem}[section]

\newtheorem{lemma}[theorem]{Lemma}

\DeclareMathSymbol{\leqslant}{\mathalpha}{AMSa}{"36}
\DeclareMathSymbol{\geqslant}{\mathalpha}{AMSa}{"3E}
\DeclareMathSymbol{\doteqdot}{\mathalpha}{AMSa}{"2B}
\DeclareMathSymbol{\circlearrowright}{\mathalpha}{AMSa}{"08}
\DeclareMathSymbol{\subsetneq}{\mathalpha}{AMSb}{"28}
\DeclareMathSymbol{\supsetneq}{\mathalpha}{AMSb}{"29}
\renewcommand{\leq}{\;\leqslant\;}
\renewcommand{\geq}{\;\geqslant\;}

\newcommand{\e}[1]{\,{\rm e}^{#1}\,}

\newcommand{\sumtwo}[2]{\sum_{\substack{#1 \\ #2}}}

\DeclareMathOperator*{\union}{\text{\large$\cup$}}
\DeclareMathOperator*{\Union}{\text{\Large$\cup$}}
\newcommand{\Uniontwo}[2]{\Union_{\substack{#1 \\ #2}}}
\DeclareMathOperator*{\inter}{\text{\large$\cap$}}

\newcommand{\expval}[1]{\langle #1 \rangle}

\newcommand{\compl}{{\text{\rm c}}}

\newcommand{\upchi}{\raise 2pt \hbox{$\chi$}}
\def\writefig#1 #2 #3 {\rlap{\kern #1 truecm \raise #2 truecm
\hbox{#3}}}

\newcommand{\caB}{{\mathcal B}}

\newcommand{\bbB}{{\mathbb B}}

\newcommand{\bbN}{{\mathbb N}}

\newcommand{\bbZ}{{\mathbb Z}}

\begin{document}

%{\hfill\small \version} \vspace{2mm}

\title{On a model of random cycles}

\author{Daniel Gandolfo, Jean Ruiz, Daniel Ueltschi}

%\iffalse
\address{Daniel Gandolfo and Jean Ruiz \hfill\newline
Centre de Physique Th\'eorique, UMR CNRS 6207 \hfill\newline
%Centre de Physique Th\'eorique, CNRS \hfill\newline
Universit\'es Aix-Marseille 1 et 2 et Sud Toulon-Var \hfill\newline
Laboratoire affili\'e \`a la FRUMAM (FR2291) \hfill\newline
Luminy, Case 907, 13288 Marseille, France
}
\email{daniel.gandolfo@cpt.univ-mrs.fr}
\email{ruiz@cpt.univ-mrs.fr}

\address{Daniel Ueltschi \hfill\newline
Department of Mathematics \hfill\newline
University of Warwick \hfill\newline
Coventry, CV4 7AL, England \hfill\newline
{\small\rm\indent http://www.ueltschi.org}
}
\email{daniel@ueltschi.org}
%\fi

\maketitle

\begin{quote}
{\small
{\bf Abstract.}
We consider a model of random permutations of the sites of the cubic lattice.
Permutations are weighted so that sites are preferably sent onto neighbors.
We present numerical evidence for the occurrence of a transition to a phase with infinite, macroscopic cycles.
}  % end \small

\vspace{1mm}

\vspace{1mm}
\noindent
{\footnotesize {\it 2000 Math.\ Subj.\ Class.:} 60K35, 82B20, 82B26, 82B41.}
\end{quote}

\section{Introduction}

Geometric representations of systems of statistical physics have a long history going back to the treatment of the Ising model by Peierls \cite{Pei}.
The Feynman-Kac formula provides such a representation for quantum models. It was originally introduced for the Bose gas \cite{Fey}, where
the symmetric nature of the particles leads to the occurrence of random permutations.
S\"ut\H o showed in the ideal gas that Bose-Einstein condensation occurs if and only if infinite cycles are present \cite{Suto1,Suto2}.
However, this relation does not seem to be always true in interacting systems \cite{Uel2}.
Spin models also have representations where correlations are represented by loops, see e.g.\ \cite{AN,CPS,Toth}.

The behavior of systems of interacting particles is notoriously difficult.
On the other hand, the presence of infinite cycles should not depend closely on the microscopic details of the model.
Results about simpler models of cycles could prove useful, especially for the understanding of the critical behavior.
The purpose of the present article is to discuss such a model.

An appropriate model of random permutations must involve the spatial nature of the original physical system.
Namely, particles are spread over a certain domain, and distant particles are not directly correlated.
The behavior of the model should also depend on the dimension of the space.
Recall that the Bose-Einstein condensation takes place in three dimensions, but not in one or two.
It seems therefore natural to consider permutations $\bbZ^d \to \bbZ^d$ instead of $\bbN \to \bbN$.
For $\Lambda \subset \bbZ^d$ finite, we will take the probability for a permutation $\pi : \Lambda \to \Lambda$ to be proportional to
\[
\prod_{x\in\Lambda} \e{-\alpha |x-\pi(x)|^2}.
\]
The motivation for these Gaussian weights comes from the Wiener measure for bosonic trajectories in the Feynman-Kac representation of the Bose gas.
The parameter $\alpha$ is proportional to the temperature of the system ({\it not} to the inverse temperature, as usual in statistical mechanics!).
This model was progressively introduced in \cite{Fey,Kik,KDS,Fic}.

There should be no long cycles when $\alpha$ is large, i.e.\ when sites are heavily discouraged from jumping to a neighbor.
Cycles should increase in size when $\alpha$ decreases.
The main question is whether a transition occurs for some value $\alpha_{\rm c}>0$, below which a fraction of the sites find themselves in infinitely long cycles.

We present numerical evidence that points towards the occurrence of infinite cycles in three dimensions.
A fraction of sites with positive density belong to infinite cycles.
A natural question is how many large cycles are found in a typical permutation, on a cube of size $L$?
One can argue that cycles are similar to a closed random walk, which has Hausdorff dimension two, so their number should grow like $L^3/L^2 = L$.
But one can also argue that infinite cycles represent correlations in the Bose condensate --- two particles are correlated if they belong to the same cycle.
In this case, there should be just one macroscopic cycle.

The surprise is that neither conclusion above is correct:
As it turns, infinite cycles are {\it macroscopic}, i.e.\ they involve a positive fraction of the sites, and their number fluctuates.
This behavior was observed in the ideal Bose gas \cite{Suto2}.

The model of random permutations is introduced in Section \ref{secmodel}, where it is also shown that there are no infinite cycles at high temperature.
A detailed probabilistic setting is described in Section \ref{secprobmodel}.
Our numerical results are presented in Section \ref{secnumresults}.
They show a surprisingly close relationship between the random cycle model and the ideal Bose gas.
The systems do not only share similar features qualitatively, but also {\it quantitatively}!

\section{The model}
\label{secmodel}

As is usual in statistical mechanics, we define the model first in a bounded domain, and then consider an appropriate thermodynamic limit.
Let $\Lambda \subset \bbZ^d$ be a large but finite cubic box centered at the origin, and let $\bbB_\Lambda$ denote the set of permutations on $\Lambda$ (i.e.\ bijections $\Lambda\to\Lambda$).
The probability for $\pi \in \bbB_\Lambda$ is given by
\be
\label{basicdef}
P_\Lambda(\pi) = \frac1{Z(\Lambda)} \prod_{x\in\Lambda} \e{-\alpha |x-\pi(x)|^2}.
\ee
Here, the parameter $\alpha$ represents the temperature, $|\cdot|$ denotes the Euclidean distance in $\bbZ^d$, and $Z(\Lambda)$ is the partition function
\be
\label{defnormalization}
Z_\Lambda = \sum_{\pi \in \bbB_\Lambda} \prod_{x\in\Lambda} \e{-\alpha |x-\pi(x)|^2}.
\ee

A {\it cycle} $\gamma$ of length $k$ is a $k$-tuple of distinct sites $(x_1,\dots,x_k)$.
Given a permutation $\pi$, we let $\gamma_0$ denote the cycle that contains the origin, and $\ell_0 = \ell_0(\pi)$ its length. That is,
\be
\gamma_0 = \bigl( 0, \pi(0), \pi^2(0), \dots, \pi^{\ell_0-1}(0) \bigr).
\ee
Notice that $1 \leq \ell_0 \leq |\Lambda|$.

The interesting phenomenon is the possible occurrence of infinite cycles, that may occur in the thermodynamic limit.
This motivates us to introduce the function $\varphi(\alpha)$ that represents the probability that the origin belongs to an infinite cycle.
First, we define
\be
\label{defPLambda}
P_\Lambda(\ell_0=k) = \sum_{\pi\in\bbB_\Lambda : \ell_0=k} P_\Lambda(\pi),
\ee
and
\be
\label{defpl}
P(\ell_0=k) = \lim_{\Lambda\nearrow\bbZ^d} P_\Lambda(\ell_0=k).
\ee
The existence of the thermodynamic limit is guaranteed by Cantor's diagonal argument:
There exists a subsequence $\Lambda_m$ of increasing cubes such that $P_{\Lambda_m}(\ell_0=k)$ converges simultaneously for all $k$.
In this paper, the notation $\Lambda\nearrow\bbZ^d$ always refers to this specific subsequence.
It is clear that $\sum_k P_\Lambda(\ell_0=k)$ is equal to one for each finite $\Lambda$.
But it may be strictly less than one in the limit $\Lambda \nearrow \bbZ^d$ (this is formalized by Fatou's lemma in analysis).
The probability $\varphi(\alpha)$ that the origin belongs to an infinite cycle is then defined by
\be
\label{defphialpha}
\varphi(\alpha) = 1 - \sum_{k\geq1} P(\ell_0=k).
\ee

Let $\alpha_{\rm c}$ denote the critical temperature,
\be
\label{defcritalpha} \alpha_{\rm c} = \sup\{ \alpha :
\varphi(\alpha)>0 \}.
\ee
We expect that $\varphi(\alpha)$ is monotone decreasing in $\alpha$ (and strictly
monotone decreasing for $\alpha<\alpha_{\rm c}$), but we are unable to prove it.
If one chooses $\alpha=0$ in \eqref{basicdef}, one easily checks that $P_\Lambda(\ell_0=k) = \frac1{|\Lambda|}$ for all $k$; then $P(\ell_0=k) = 0$, and $\varphi(0) = 1$.

It is easy to show the absence of infinite cycles when $\alpha$ is large enough.
The following theorem implies that $\alpha_{\rm c} < \infty$ for any dimension.

\begin{theorem}
\label{thmnoinfinitecycles}
Assume that
$$
\sum_{x \in \bbZ^d \setminus \{0\}} \e{-\alpha |x|^2} < 1.
$$
Then $\varphi(\alpha)=0$.
\end{theorem}

\begin{proof}
Given a cycle $\gamma = (x_1,\dots,x_k)$, let
\be
\omega(\gamma) = \prod_{j=1}^k \e{-\alpha |x_{j+1}-x_j|^2}
\ee
(we set $x_{k+1}=x_1$).
From the definition \eqref{basicdef}, we have
\be
P_\Lambda(\ell_0=k) = \sum_{\gamma = (0,x_2,\dots,x_k)} \omega(\gamma) \frac{Z(\Lambda \setminus \{0,x_2,\dots,x_k\})}{Z(\Lambda)}.
\ee
The ratio of partition functions is less than one.
Since $\omega(\gamma) \leq \prod_{j=1}^{k-1} \e{-\alpha |x_{j+1}-x_j|^2}$, and neglecting the restriction that cycles do not self-intersect, we get
\be
\label{bornePLambda}
P_\Lambda(\ell_0=k) \leq \Bigl( \sum_{x \in \bbZ^d \setminus \{0\}} \e{-\alpha |x|^2} \Bigr)^{k-1}.
\ee
Then $P_\Lambda(\ell_0=k) \leq c^{k-1}$ with $c<1$ for all domains $\Lambda$.
From the definitions \eqref{defPLambda}--\eqref{defphialpha}, we have
\be
\begin{split}
\varphi(\alpha) &= \lim_{N\to\infty} \lim_{\Lambda\nearrow\bbZ^d} 1 - \sum_{k=1}^N P_\Lambda(\ell_0=k) \\
&= \lim_{N\to\infty} \lim_{\Lambda\nearrow\bbZ^d} \sum_{k=N+1}^{|\Lambda|} P_\Lambda(\ell_0=k).
\end{split}
\ee
The sum is bounded by a convergent geometric series uniformly in $\Lambda$, see \eqref{bornePLambda}, which implies that $\varphi(\alpha)=0$.
\end{proof}

We can define a thermodynamic potential by
\be
\label{defthermopot}
f(\alpha) = \lim_{\Lambda \nearrow \bbZ^d} \frac1{|\Lambda|} \log Z_\Lambda.
\ee
Here, we can take the limit along any sequence of boxes of increasing sizes, as can be established by a standard subadditive argument.
The function $f(\alpha)$ is convex, since $\frac{\partial^2}{\partial\alpha^2} \log Z_\Lambda$ is given by the expectation of positive fluctuations.
We conjecture that $f(\alpha)$ is analytic for all $\alpha$, except for $\alpha=\alpha_{\rm c}$.

\section{The probability model}
\label{secprobmodel}

This section is more technical and it can be skipped on first reading.
But we hope to catch the interest of probabilists, which prompts us to turn the ideas above into a probability model.
The situation is actually not trivial; our results require a condition, see \eqref{assumption} below, that we cannot prove in the case of Gaussian weights.
The elements of the probability space are lattice permutations.
The $\sigma$-algebra contains the event that the origin belongs to an infinite cycle, and the probability measure is inspired by \eqref{basicdef}.

\subsection{The probability space}

Let $\bbB$ be the set of all permutations $\pi$ on $\bbZ^d$ (i.e.\ bijections $\bbZ^d \to \bbZ^d$).
We redefine $\bbB_\Lambda$ so that it is a subset of $\bbB$, by setting
\be
\bbB_\Lambda = \bigl\{ \pi \in \bbB : \pi(x)=x \text{ for all } x \notin \Lambda \bigr\}.
\ee
Let $B_{xy}$ be the set of permutations such that $x$ is sent onto $y$,
\be
B_{xy} = \bigl\{ \pi \in \bbB : \pi(x)=y \bigr\}.
\ee
We let $\caB$ denote the $\sigma$-algebra generated by $\{B_{xy}\}_{x,y \in \bbZ^d}$.
For any $n$ and any $x_1,\dots,x_n$, $y_1,\dots,y_n \in \bbZ^d$, the probability of the set
\be
\label{intersets}
B = \inter_{i=1}^n B_{x_i y_i}
\ee
is defined by
\be
\label{defmeasure}
p(B) = \lim_{\Lambda\nearrow\bbZ^d} \frac1{Z(\Lambda)} \sum_{\pi \in B \cap \bbB_\Lambda} \prod_{x\in\Lambda} \e{-\alpha \, \xi(x,\pi(x))}.
\ee
Here, $\xi(x,y)$ is a nonnegative, symmetric function on $\bbZ^d \times \bbZ^d$.
An interesting example is $\xi(x,y) = |x-y|^2$.
The normalization $Z(\Lambda)$ is given by
\be
Z(\Lambda) = \sum_{\pi \in \bbB_\Lambda} \prod_{x\in\Lambda} \e{-\alpha \, \xi(x,\pi(x))}.
\ee
The thermodynamic limit $\Lambda\nearrow\bbZ^d$ in \eqref{defmeasure} exists at least on a subsequence of increasing cubes, thanks again to Cantor's diagonal process.

We now introduce the event that the origin belongs to an infinite cycle.
First, consider
\be
B_0^{(k)} = \union_{x_2,\dots,x_k} \inter_{i=1}^k B_{x_i x_{i+1}},
\ee
where $x_1 = x_{k+1} = 0$. The union is over distinct sites $x_2, \dots, x_k \in \bbZ^d$ (they are also distinct from 0).
The set $B_0^{(k)}$ represents the event that the origin belongs to a cycle of length $k$.
Next, let
\be
B_0^{(\infty)} = \Bigl[ \union_{k\geq1} B_0^{(k)} \Bigr]^\compl
\ee
be the event for the origin to belong to a cycle of infinite length.
It is clear that $B_0^{(\infty)}$ belongs to the $\sigma$-algebra $\caB$.

The proof of the existence of a measure is not trivial, and we need the following assumption: For any $x \in \bbZ^d$, we suppose that
\be
\label{assumption}
\sum_{y \in \bbZ^d} p(B_{xy}) = 1.
\ee
This is equivalent to another condition that is more technical, but also more explicit:
\be
\label{assumptionbis}
\lim_{R\to\infty} \lim_{\Lambda \nearrow \bbZ^d} \sumtwo{y \in \bbZ^d}{|y-x|>R} p_\Lambda(B_{xy}) = 0.
\ee
This condition means that sites do not jump straight to infinity in one step.
Notice that it fails to be true when $\alpha=0$.
It trivially holds if $\xi(x,y)=\infty$ when $|x-y|$ is larger than some cutoff distance.
It can also be established for $\xi(x,y) = |x-y|^2$ when $\alpha$ is large, but we cannot prove it for arbitrary $\alpha>0$, although it is certainly true.

\begin{theorem}
\label{thmexistencemeasure}
Assume that \eqref{assumption} holds true.
Then there exists a unique probability measure $p$ on $(\bbB,\caB)$ that coincides with \eqref{defmeasure} for all sets of the form \eqref{intersets}.
In addition, we have $p(B_0^{(\infty)}) = \varphi(\alpha)$, where $\varphi(\alpha)$ is defined as in Section \ref{secmodel}, but with $\xi(x,y)$ instead of $|x-y|^2$.
\end{theorem}

This theorem is proved at the end of Section \ref{secconstrmeasure}.
The condition \eqref{assumption} is necessary.
Otherwise, there exists $x \in \bbZ^d$ such that $\sum_y p(B_{xy}) = 1-\varepsilon$ with $\varepsilon>0$.
Consider then the decreasing sequence of sets $(B_n)$ with $B_n = \cup_{y, |y-x|>n} B_{xy}$.
Then $\cap_n B_n = \emptyset$, but
\be
p(B_n) = 1 - \sum_{y, |y-n|>n} p(B_{xy}) \geq \varepsilon,
\ee
so $p(B_n)$ does not go to 0 and $p$ cannot be a probability measure.

Notice that Theorem \ref{thmnoinfinitecycles}, in Section \ref{secmodel}, extends straightforwardly to general $\xi(x,y)$.
It follows that $P(B_0^{(\infty)})=0$ when $\alpha$ satisfies $\sum_{x\neq0} \e{-\alpha \, \xi(0,x)} < 1$.

\subsection{Construction of the measure}
\label{secconstrmeasure}

Let us introduce the algebra $\caB'$ generated by the sets $B_{xy}$.
It is not hard to verify that the set function defined in \eqref{defmeasure} extends uniquely to a finitely additive measure on $\caB'$.
We need to prove that it is $\sigma$-additive within the algebra.

\begin{lemma}
\label{lemsigmaadditivity}
Suppose that the property \eqref{assumption} is satisfied.
Let $B_1 \supset B_2 \supset \dots$ be any sequence of sets of $\caB'$ that decreases to $\emptyset$, i.e.\ $\inter_{n\geq1} B_n = \emptyset$.
Then $\lim_{n\to\infty} p(B_n) = 0$.
\end{lemma}

\begin{proof}
We show the counterpositive, namely that if $(B_n)$ is a decreasing sequence in $\caB'$ such that $p(B_n) > \varepsilon > 0$ for any $n$, then $\cap_n B_n$ is not empty.

Given $\pi\in\bbB$ and $\Lambda \subset \bbZ^d$, we denote $A(\pi;\Lambda)$ the set of permutations whose restriction to $\Lambda$ coincides with $\pi$.
Precisely,
\be
A(\pi;\Lambda) = \bigl\{ \pi' \in \bbB : \pi'(x) = \pi(x) \text{ for all } x \in \Lambda \bigr\}.
\ee
Since $B_n \in \caB'$, for each $n$ there exists a finite set $\Lambda_n$ such that $B_n$ is given by
\be
\label{disguisedcountableunion}
B_n = \union_{\pi \in B_n} A(\pi;\Lambda_n).
\ee
The set $B_n$ is uncountable, but there are only countably many {\it distinct} sets $A(\pi;\Lambda_n)$.
Thus Eq.\ \eqref{disguisedcountableunion} is really a countable union of disjoint sets.
We can suppose that $(\Lambda_n)$ is an increasing sequence of cubes centered at the origin.
Let $k_n$ be some integer, and $C_n \subset B_n$ be the set of permutations where each site of $\Lambda_n$ is sent at distance less than $k_n$:
\be
C_n = \bigl\{ \pi \in B_n : |\pi(x)-x| \leq k_n \text{ for all } x \in \Lambda_n \bigr\}.
\ee
We have
\be
B_n \setminus C_n \subset \Union_{x\in\Lambda_n} \Uniontwo{y \in \bbZ^d}{|y-x|>k_n} B_{xy}.
\ee
It follows that
\be
\begin{split}
p(B_n \setminus C_n) &= \lim_{\Lambda \nearrow \bbZ^d} p_\Lambda(B_n \setminus C_n) \\
&\leq \sum_{x\in\Lambda_n} \lim_{\Lambda \nearrow \bbZ^d} \sum_{y, |y-x|>k_n} p_\Lambda(B_{xy}).
\end{split}
\ee
From the assumption \eqref{assumptionbis} we can choose $k_n$ large enough so that $p(B_n \setminus C_n) \leq \frac12 \varepsilon$.
Then $(C_n)$ is a decreasing sequence such that $p(C_n) \geq \frac12 \varepsilon$ for all $n$.

In a fashion similar to \eqref{disguisedcountableunion}, we can decompose $C_n$ into the disjoint union
\be
\label{finiteunion}
C_n = \Union_{i=1}^{r_n} A(\pi_{n,i};\Lambda_n)
\ee
for some suitable permutations $\pi_{n,i}$.
The number $r_n$ is now finite.
We have $\Lambda_n \subset \Lambda_{n+1}$.
Two sets $A(\pi,\Lambda_n)$ and $A(\pi',\Lambda_{n+1})$ are either disjoint, or $A(\pi,\Lambda_n) \supset A(\pi',\Lambda_{n+1})$ if the restrictions of $\pi$ and $\pi'$ on $\Lambda_n$ coincide.
We can define a tree with vertices $(n,i)$, $n=1,2,\dots$ and $1\leq i\leq r_n$, and with an edge between $(n,i)$ and $(n+1,j)$ whenever
\be
A(\pi_{n,i},\Lambda_n) \supset A(\pi_{n+1,j},\Lambda_{n+1}).
\ee
We also connect the vertices $(1,i)$, $1\leq i\leq r_1$, to the root 0.
There are $r_n$ vertices at distance $n$ from the root, so all incidence numbers are finite.
For each vertex $(n,i)$, let $q(n,i)$ denote the length of the longest path descending from $(n,i)$.
For each $n$ it is infinite for at least one $i$.
In addition, each $(n,i)$ with $q(n,i)=\infty$ is connected to at least one vertex $(n+1,j)$ with $q(n+1,j)=\infty$.
These properties hold true because incidence numbers are finite, i.e.\ because we defined the tree starting with the sequence $(C_n)$ instead of $(B_n)$.
We can select an infinite path $(n,j_n)$ with $q(n,j_n)=\infty$ for all $n$.
Since any $x$ belongs to some $\Lambda_n$, we can define a map $\pi : \bbZ^d \to \bbZ^d$ by setting
\be
\pi(x) = \pi_{n,j_n}(x).
\ee
Observing that $\pi$ is a permutation, and that $\pi \in C_n$ for all $n$, we see that $\cap_n C_n$ is not empty.
\end{proof}

\begin{proof}[Proof of Theorem \ref{thmexistencemeasure}]
It follows from Lemma \ref{lemsigmaadditivity} that $p$ is a $\sigma$-additive premeasure on $\caB'$.
It has a unique extension to a measure on $\caB$ by the Carath\'eodory-Fr\'echet extension theorem.

In order to show that $p(B_0^{(\infty)})$ is equal to the probability of infinite cycles as defined in \eqref{defphialpha}, let us observe that
$B_0^{(k)} \cap B_0^{(k')} = \emptyset$ if $k \neq k'$, and $B_0^{(\infty)} = [\cup_k B_0^{(k)}]^\compl$.
One can check that $P(\ell_0=k)$ defined in \eqref{defpl} is equal to $p(B_0^{(k)})$.
The result then follows from the $\sigma$-additivity of $p$.
\end{proof}

\section{Numerical results}
\label{secnumresults}

\subsection{Description of the method}

We have performed intensive Monte Carlo simulations of the random cycle model with Gaussian weights for the jumps.
The dynamics is pure Metropolis; a change in the permutation configuration is accepted or rejected according to the change in ``energy''.

A ``step'' of our code consists in sweeping the sites in lexicographic order.
For each site $x$, we randomly choose a site $y$ in a window centered at $x$ with size depending on the temperature $\alpha$.
Given $x,y$, we consider changing the permutation $\pi$ into $\pi'$, where $\pi'$ is defined as follows (see Fig.\ \ref{figmcupdate}):
\be
\begin{cases} 
\pi'(x) = y, \\
\pi' \bigl( \pi^{-1}(y) \bigr) = \pi(x),
\end{cases}
\ee
and $\pi'(z) = \pi(z)$ for $z \neq x, \pi^{-1}(y)$.

\bfig
\setlength{\unitlength}{0.2in}
\centering \begin{picture}(13,11)(0,-17)
\put(12,-7.5){\Large{$y$}}
\put(1.5,-7.5){\Large{$x$}}
\put(0.5,-16){\Large{$\pi(x)$}}
\put(11.5,-16){\Large{$\pi^{-1}(y)$}}
\put(11,-8){\circle*{0.5}}
\put(3,-8){\circle*{0.5}}
\drawline(3,-8)(3,-15)
\dashline{0.30}(3,-8)(11,-8)
\put(6.8,-8.22){$\blacktriangleright$}
\put(2.72,-11.6){$\blacktriangledown$}
\put(11,-15){\circle*{0.5}}
\put(3,-15){\circle*{0.5}}
\drawline(11,-8)(11,-15)
\put(6.8,-15.24){$\blacktriangleleft$}
\dashline{0.30}(3,-15)(11,-15)
\put(10.72,-11.6){$\blacktriangle$}
\end{picture}
\caption{Illustration for the update scheme of the algorithm.
Bold lines represent the old permutation $\pi$, dashed lines represent the new permutation $\pi'$.}
\label{figmcupdate}
\efig

The energy difference between new and old permutations is $\Delta H = H_\Lambda(\pi') - H_\Lambda(\pi)$, where 
$H_\Lambda(\pi) = \sum_{x\in\Lambda} |x-\pi(x)|^2$ denotes the ``Hamiltonian'' of the model.
We have
\be
\Delta H = \bigl| x - y \bigr|^2 + \bigl| \pi^{-1}(y) - \pi(x) \bigr|^2 - \bigl| x - \pi(x) \bigr|^2 - \bigl| \pi^{-1}(y) - y \bigr|^2.
\ee
Then according to the Metropolis prescription, the change is accepted with probability
\[
\min[1,\e{-\alpha \Delta H}].
\]

The initial configuration is usually the identity permutation $\pi(x) \equiv x$, but we have also considered initial random configurations chosen over all permutations $\Lambda \to \Lambda$ with uniform distribution.
It turns out that the system thermalizes extremely well, irrespective of the initial condition.
Measurements are taken after suitable thermalization.
All numerical computations were performed on a personal computer and they can be reproduced rather easily.

Most of our measurements are for the random variable $\rho_k$, that represents the fraction of sites that belong to cycles of length {\it less than or equal to} $k$.
Precisely, $\rho_k$ is defined by
\be
\rho_k(\pi) = \frac{\#\{ x\in\Lambda : \ell_x(\pi) \leq k \}}{|\Lambda|},
\ee
where $\ell_x$ is the length of the cycle that contains $x$. We have $0 \leq \rho_k(\pi) \leq 1$ and $1 \leq k \leq |\Lambda|$.
It is related to $\varphi(\alpha)$ if we assume that the system is essentially translation invariant.
We expect that
\be
\varphi(\alpha) \approx 1 - \expval{\rho_N},
\ee
with $N$ such that $1 \ll N \ll |\Lambda|$.

\subsection{Fixed temperature, different dimensions}

We have first fixed $\alpha=0.2$ and considered cubic boxes in dimensions $d=1$, $d=2$, and $d=3$.
Fig.\ \ref{figallDsmallT} depicts the graphs of the expectation $\expval{\rho_k}$ of the fraction of sites in cycles of length less than $k$.
The horizontal axis in Fig.\ \ref{figallDsmallT} are $k/|\Lambda|$; they take values between 0 and 1.
\bfig
\centerline{\includegraphics[width=120mm]{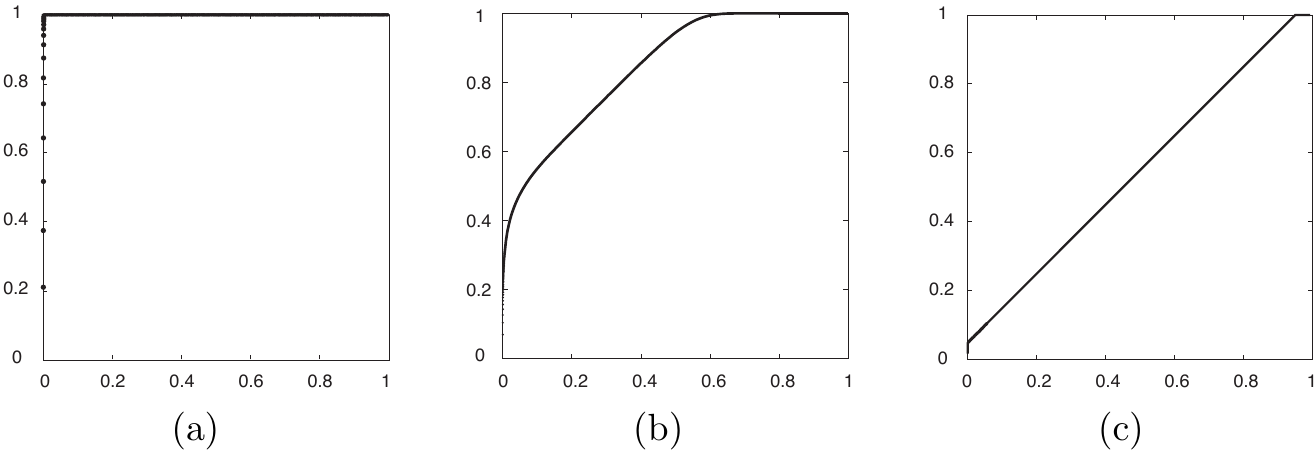}}
%\centerline{\epsfxsize=120mm \epsffile{figallDsmallT.eps}}
\caption{Expectation value for $\rho_k$ as function of $k/|\Lambda|$ for $\alpha=0.2$.
(a) $d=1$ and $|\Lambda| = 10'000$. (b) $d=2$ and $|\Lambda| = 100^2$. (c) $d=3$ and $|\Lambda| = 50^3$.}
\label{figallDsmallT}
\efig

Fig.\ \ref{figallDsmallT} (a) shows that almost all sites belong to cycles with very small length compared to the volume of the system.
The situation is different in $d=2$ and $d=3$.
In $d=2$, around 25\% of the sites belong to small cycles, and 75\% belong to {\it macroscopic} cycles, i.e.\ to cycles whose length is a fraction of the volume.
We expect that this density is equal to $\varphi(\alpha)$ defined in \eqref{defphialpha}.
The same holds for $d=3$, with respective densities 3\% and 97\%.
Fig.\ \ref{figallDbigT} shows the situation at the higher temperature $\alpha=2$, where there are clearly no macroscopic cycles.

\bfig
\centerline{\includegraphics[width=120mm]{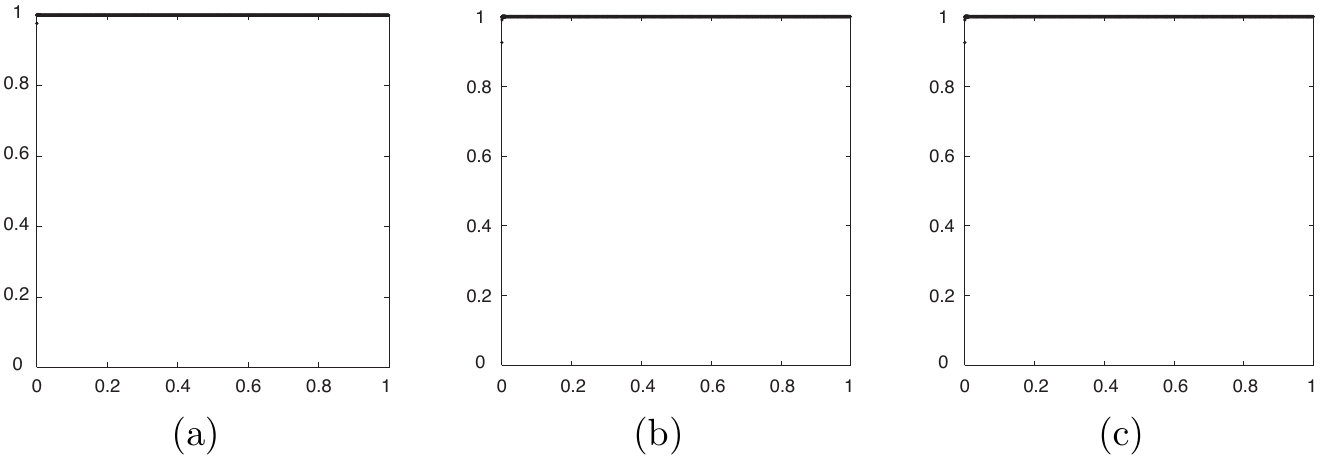}}
%\epsfxsize=120mm 
%\centerline{\epsffile{figallDbigT.eps}}
\caption{Expectation value for $\rho_k$ as function of $k/|\Lambda|$ for $\alpha=2$.
(a) $d=1$ and $|\Lambda| = 10'000$. (b) $d=2$ and $|\Lambda| = 100^2$. (c) $d=3$ and $|\Lambda| = 50^3$.}
\label{figallDbigT}
\efig

From these preliminary exploration, it seems that macroscopic cycles are present in dimensions greater or equal to 2, instead of 3!

\subsection{Two dimensions}

The main numerical result for $d=2$ is shown in Fig.\ \ref{fig2Dallsizes}, where the expectation $\expval{\rho_k}$ is plotted for $\alpha=0.1$ and different lattice sizes.
\bfig
\centerline{\includegraphics[width=100mm]{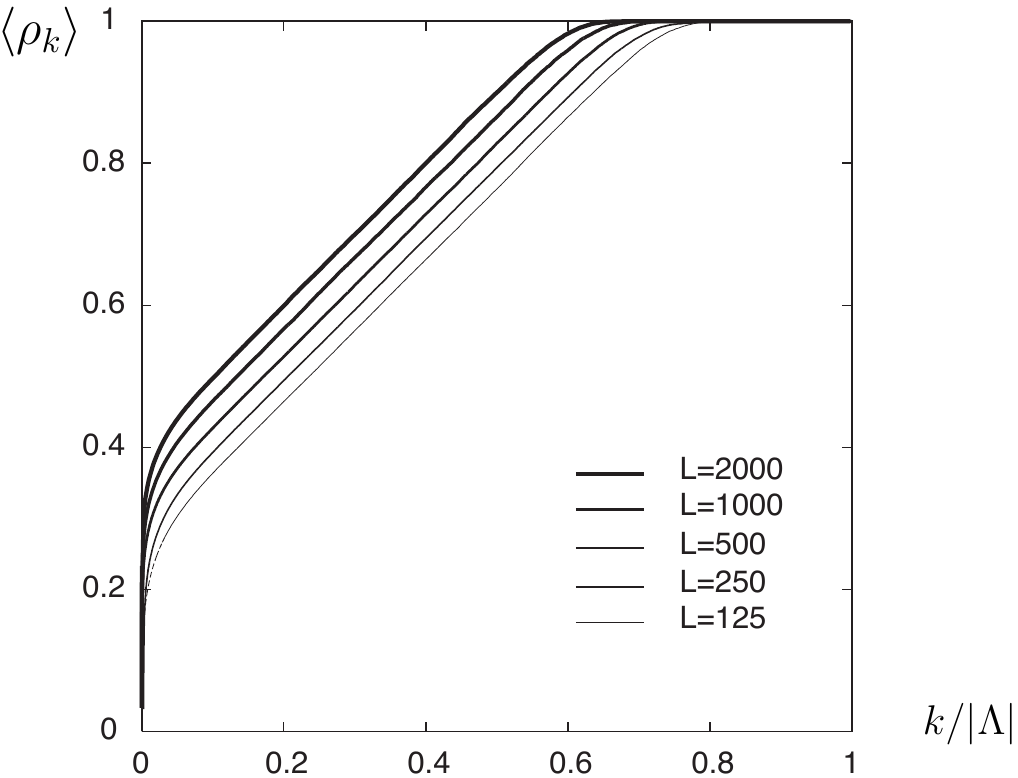}}
\caption{Expectation $\expval{\rho_k}$ for $d=2$, $\alpha=0.1$, and with various domain sizes.}
\label{fig2Dallsizes}
\efig
It is manifest that finite size effects are very important.
The density of sites in small cycles grows from 10\% for $L=50$ to 30\% for $L=2000$.
We expect the curves to continue their progression upwards as the size increases, until no macroscopic cycles is left in the limit $L\to\infty$.

A possible explanation involves random walks.
The cycle containing the origin is a self-avoiding closed random walk --- but its probability differs from that of random walks because of the presence of all other cycles.
Nonetheless, the analogy with random walks is worth pursuing.
Random walks are recurrent in $d=2$.
The probability $f_n$ for the simple random walk to return to the origin for the first time after $n$ steps satisfies \cite{BF}
\be
\sum_{n\geq1} f_n = 1, \quad\quad f_n \sim (n \log^2 n)^{-1}.
\ee
The ``macroscopic cycles'' in $d=2$ are those that are big with respect to the size of the domain, i.e.\ whose mean square distance $\sqrt n$ is larger than the size $L$ of the domain.
Let $\rho(L)$ denote the density of sites in long cycles.
The condition $\sqrt n \sim L$ implies that the probability for the origin to belong to a long cycle is roughly
\be
\rho(L) \sim \sum_{n \geq L^2} f_n \sim (\log L)^{-1}.
\ee
It follows that
\be
\rho(L) - \rho(2^k L) \sim \frac k{\log^2 L} \, \frac1{1 + k \frac{\log 2}{\log L}}.
\ee
In first approximation this is proportional to $k$, as observed in Fig.\ \ref{fig2Dallsizes}.

To summarize this section about two dimensions, we have understood that there are no really macroscopic cycles, and that the evidence suggested in Fig.\ \ref{figallDsmallT} (b) comes from strong finite-size effects.

\subsection{Three dimensions}

Fig.\ \ref{fig3Dallsizes} shows $\expval{\rho_k}$ for $\alpha=0.8$ and different sizes of the domain.
In contrast to Fig.\ \ref{fig2Dallsizes} there are no noticeable finite-size effects.
It is thus clear that macroscopic cycles are present in three dimensions, and that numerical simulations work remarkably well.

\bfig
\centerline{\includegraphics[width=100mm]{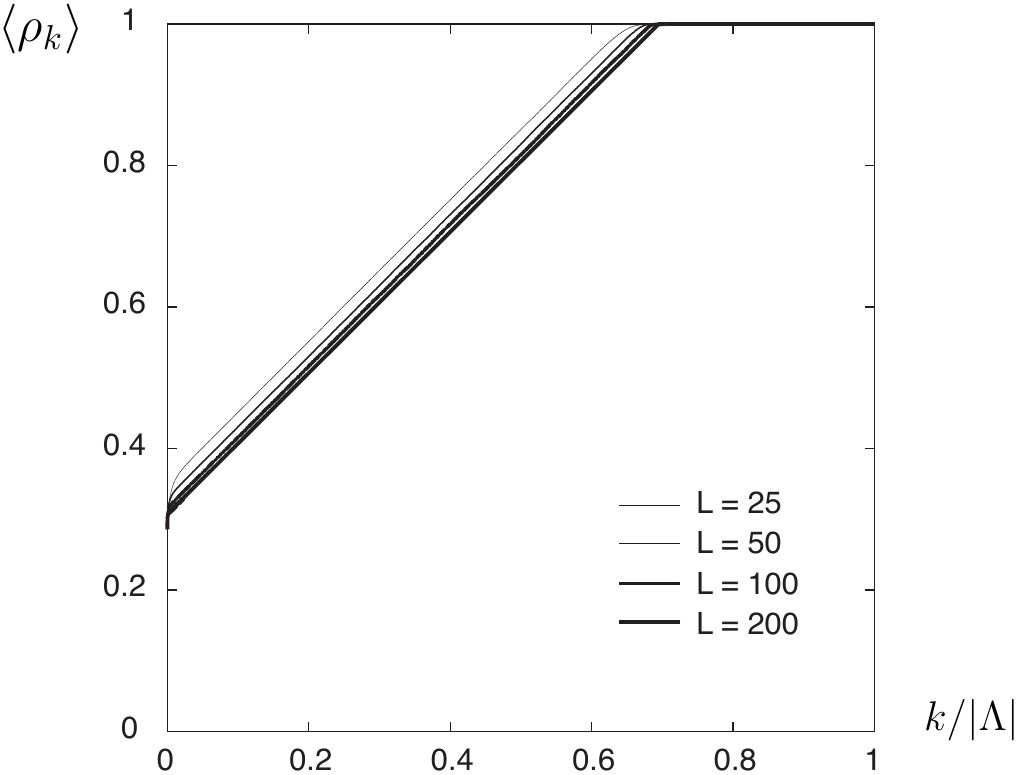}}
\caption{Expectation $\expval{\rho_k}$ for $d=3$, $\alpha=0.8$, and with various domain sizes.
We observe that $\varphi(0.8) \approx 0.7$.}
\label{fig3Dallsizes}
\efig

We have computed the density of sites in macroscopic cycles as a function of the temperature, see Fig.\ \ref{fig3Dphialpha}.
Recall that it should be equal to $\varphi(\alpha)$.
We find that $\varphi(\alpha)$ seems to be continuous, and that the critical temperature is $\alpha \approx 1.7$.
It is instructive to compare it with the critical temperature for the Bose-Einstein condensation of the ideal gas, which is known exactly.
The difference between our model and the Feynman-Kac representation of the ideal gas is that our particles are frozen on the sites of the cubic lattice.
But the comparison is otherwise possible.
\bfig
\centerline{\includegraphics[width=80mm]{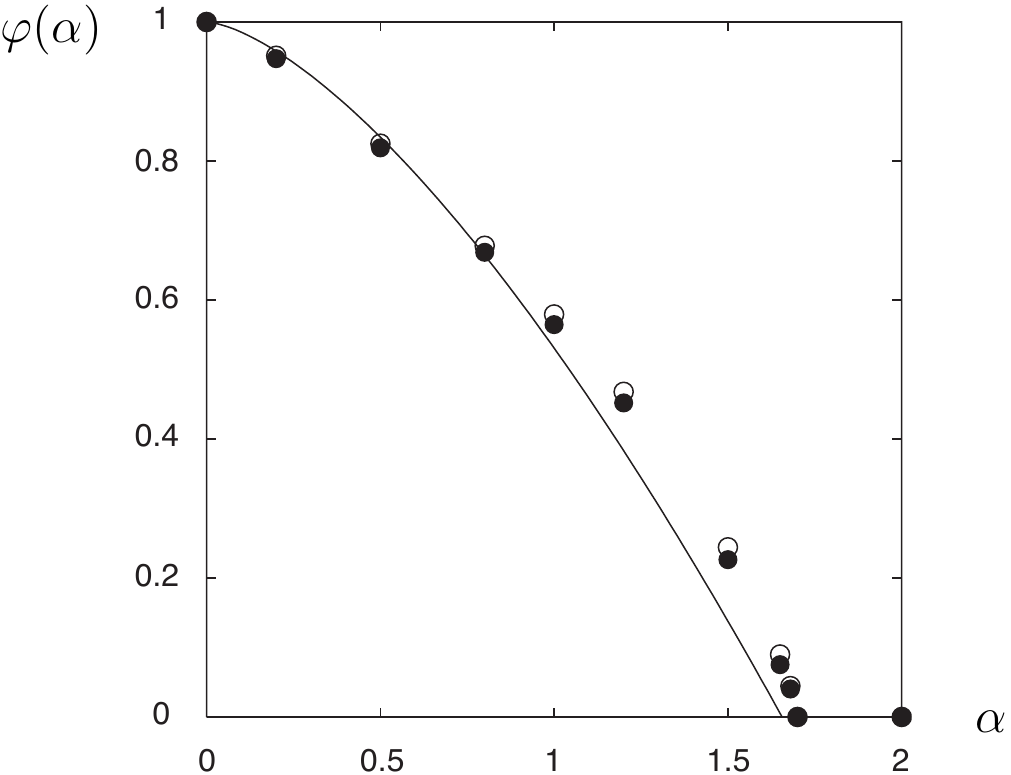}}
\caption{Density of sites in macroscopic cycles versus temperature, in three dimensions.
The points represent numerical data for $L=50$ (in black) and $L=100$ (in white).
We observe that the critical temperature of our model is $\alpha_{\rm c} \approx 1.7$.
The curve gives the density of the Bose condensate $\rho_0^{\rm BEC}(\alpha)$, Eq.\ \eqref{Bosecondensate}.}
\label{fig3Dphialpha}
\efig
The particle density in our system is equal to one, and the physical constants are $\frac\hbar{2m} = k_{\rm B} = 1$.
Then our $\alpha$ is proportional to the temperature, with $\alpha = 1/4\beta$.
With $\zeta$ denoting the Riemann zeta function, the critical temperature of the corresponding ideal Bose gas is
\be
\alpha_{\rm c}^{\rm BEC} = \zeta(3/2)^{-2/3} \pi = 1.656...
\ee
It is tantalizingly close to the critical temperature in the random cycle model!
In addition, the density of the condensate in the Bose gas is given by
\be
\label{Bosecondensate}
\rho_0^{\rm BEC}(\alpha) = 1 - \Bigl( \frac\alpha{\alpha^{\rm BEC}_{\rm c}} \Bigr)^{3/2}.
\ee
It is plotted in Fig.\ \ref{fig3Dphialpha} along $\varphi(\alpha)$.
$\varphi(\alpha)$ and $\rho_0^{\rm BEC}(\alpha)$ appear to be close, but it seems that their differences cannot be accounted for by numerical errors of by finite size effects.

All numerical results so far were for the average $\expval{\rho_k}$ over many permutations.
We now discuss the distribution of macroscopic cycles in a typical permutation.
\bfig
\centerline{\includegraphics[width=120mm]{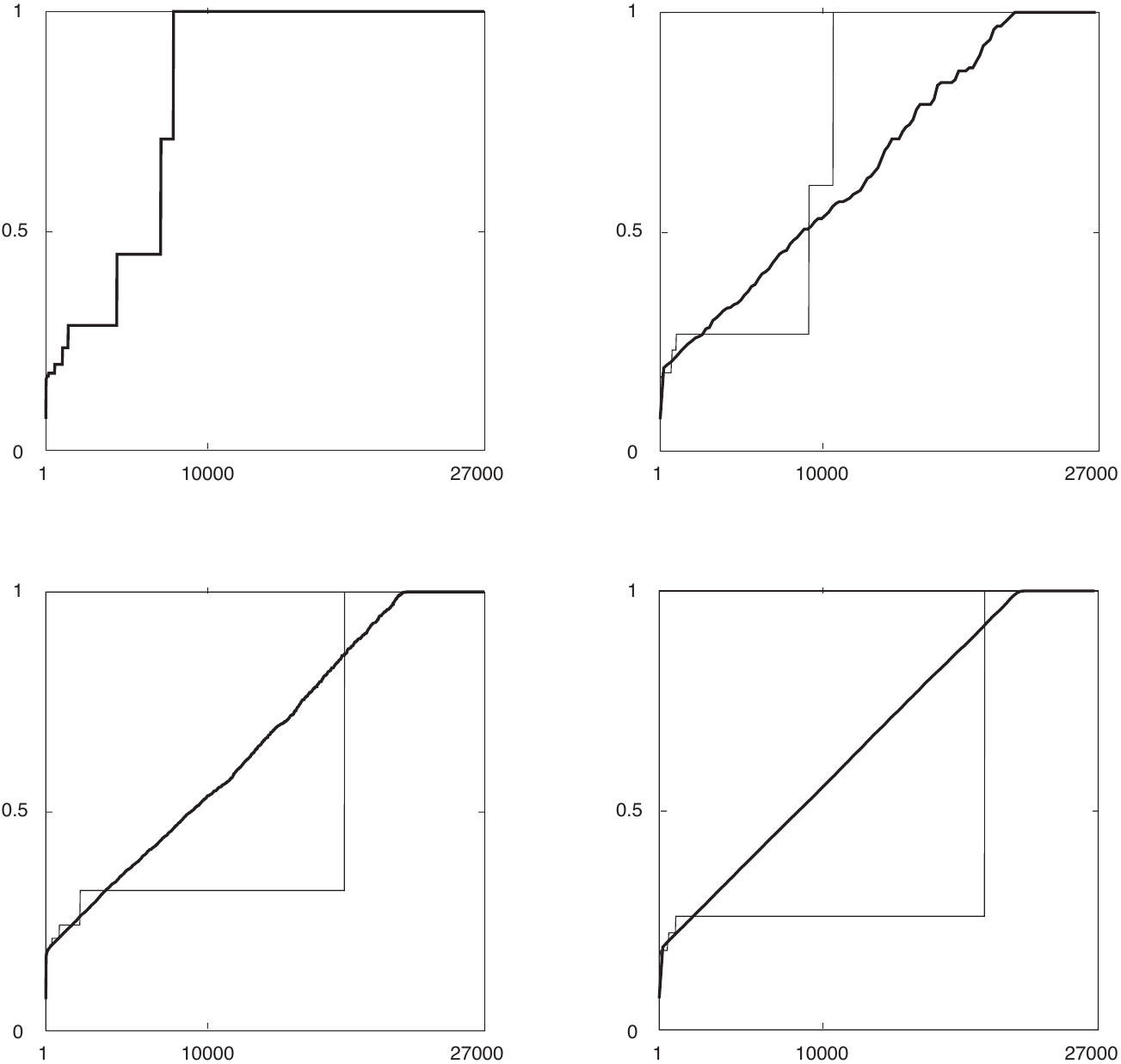}}
\caption{Thin lines show $\rho_k(\pi)$ in the permutation after 1, 100, 1000, and 20'000 steps.
The system has been thermalized initially.
Bold lines show the average $\expval{\rho_k}$ in all permutations until 1, 100, 1000, and 20'000 steps.
Here, $d=3$, $\alpha=0.5$, and $L=30$.}
\label{fig3Dtypperm}
\efig
Fig.\ \ref{fig3Dtypperm} shows $\rho_k$ for $\alpha=0.5$ and $L=30$.
We plot the value of $\rho_k(\pi)$ for the permutation $\pi$ after $\tau$ steps, and its average $\expval{\rho_k}$ over all permutations before $\tau$ steps.
The latter converges as $\tau\to\infty$, as expected.
The jumps in the graph of $\rho_k(\pi)$ correspond to macroscopic cycles.
A jump at $\frac k{|\Lambda|}$ means that a macroscopic cycle of density $\frac k{|\Lambda|}$ is present, and accordingly the height of the jump is given by
\be
\rho_k(\pi) - \rho_{k-1}(\pi) = \tfrac k{|\Lambda|}.
\ee

The number of macroscopic cycles (of density larger than $\varepsilon>0$) seems to fluctuate.
As $\Lambda \nearrow \bbZ^3$, the distribution of this number converges to some nontrivial distribution.
This behavior was observed by S\"ut\H o in the ideal gas \cite{Suto2}.
Macroscopic cycles are due to particles in the condensate, for which the underlying distribution of permutations is uniform.
Let us tentatively extrapolate this observation to our model.
It suggests that the domain $\Lambda$ splits into two sets $\Lambda_{\rm f}$ and $\Lambda_{\infty}$, that corresponds to sites in finite and macroscopic cycles, respectively.
While these sets are random, their respective volumes are close to $(1-\varphi(\alpha)) |\Lambda|$ and $\varphi(\alpha) |\Lambda|$.
Suppose that the distribution of macroscopic cycles in $\Lambda_\infty$ is the same as if the permutation on $\Lambda_\infty$ was chosen with uniform probability.
The average density of the longest cycle in uniformly distributed random permutations is known, see e.g.\ \cite{SL}, and is equal to 0.6243\dots.
We report in Table \ref{table1} the results of numerical measurements in our model.
We find values that seem to be in total agreement.
This is extremely surprising, since the jumps in our permutations satisfy certain spatial restrictions; with uniform permutations, sites can go to infinity in but one step.

\begin{table}
\begin{center}
\begin{tabular}{|c|c|c|c|}
  \quad $\alpha$ \quad & \quad $\expval{\frac{\ell_{\max}}{|\Lambda|}}$ \quad & \quad $\varphi(\alpha)$ \quad & $\expval{\frac{\ell_{\max}}{|\Lambda|}}/\varphi(\alpha)$ \\
  \hline
  0.001  &  0.6242 & 0.9998 & 0.6243 \\
  0.25   &  0.5804 & 0.9295 & 0.6244 \\
  0.50   &  0.5115 & 0.8187 & 0.6247 \\
  0.75   &  0.4338 & 0.6944 & 0.6248 \\
  1.50   &  0.1414 & 0.2264 & 0.6244 \\
  1.68   &  0.0251 & 0.0402 & 0.6258 \\
\end{tabular}
\bigskip
\caption{Average density of the longest cycle as a function of the temperature. Here, $d=3$ and $L=50$.} 
\label{table1}
\end{center}
\end{table}

\section{Conclusion}

We have considered a model of random permutations on the cubic lattice.
It is {\it a priori} a crude approximation for the ideal Bose gas in the Feynman-Kac representation.
Surprisingly, its behavior is close to the Bose gas both qualitatively and quantitatively.
The critical temperatures are very close, and the average densities of the longest cycle coincide.
But there seems to be a small difference between the density of sites in macroscopic cycles and the density of the Bose-Einstein condensate.
The study of larger systems on powerful computers may shed more light on this issue.

The fact that this simple model is so close to the ideal Bose gas suggests to use it in order to simulate interacting systems.
A natural generalization is to introduce interactions between permutation jumps.
Can we do it so that the critical temperature of the random cycle model is in quantitative agreement with the critical temperature of the interacting Bose gas?

Many mathematical aspects need clarifying as well.
One would like to know about analytic properties of the thermodynamic potential \eqref{defthermopot}.
The existence of a probability measure on the space of permutations on $\bbZ^d$ should be completed.
The absence of cycles in one and two dimensions should be established, and also certain properties such as the monotone decreasing behavior of $\varphi(\alpha)$ with respect to $\alpha$.

\bigskip
{\bf Acknowledgments.}
We are grateful to Stefan Adams, Volker Betz, Roman Koteck\'y, and Valeriy Slastikov for useful discussions, and to Franco Vivaldi for informing us of the reference \cite{SL}.
D.\ G.\ and J.\ R.\ thank the University of Warwick, and D.\ U.\ thanks the Centre de Physique Th\'eorique in Marseille, for kind hospitality.
This research was supported in part by the grant DMS-0601075 of the US National Science Foundation, and by the grant RD06020 of the University of Warwick.

\end{document}